\documentclass[preprint,12pt]{elsarticle}

\usepackage{amssymb}
\usepackage{soul}
\usepackage{comment}
\usepackage{lineno}

  \usepackage[hidelinks]{hyperref}

\usepackage[dvipsnames]{xcolor}

\begin{document}

\begin{frontmatter}

\title{Characterizing the Experiment for Calibration \newline with Uranium (Excalibur) Neutron Source \newline for Use in Warhead Verification}

\author[PU]{Jihye~Jeon}
\author[PPPL]{Erik~P.~Gilson}
\author{Michael~Hepler}
\author[PU]{Alexander~Glaser}
\author[PU,PPPL]{Robert~J.~Goldston}

\affiliation[PU]{organization={Princeton University},
            addressline={221 Nassau St., 2nd Fl.}, 
            city={Princeton},
            postcode={08542}, 
            state={NJ},
            country={USA}}

\affiliation[PPPL]{organization={Princeton Plasma Physics Laboratory},
            addressline={100 Stellarator Rd.}, 
            city={Princeton},
            postcode={08540}, 
            state={NJ},
            country={USA}}

\begin{abstract}

Neutron sources can play a variety of roles in warhead verification. For transmission radiography, a source of directed high energy neutrons is required, while for applications to detect fissile isotopes, sub-MeV neutrons are preferred. The Excalibur (Experiment for Calibration with Uranium) neutron source has been built and used in a variety of verification-related experiments. Excalibur is based on a commercial deuterium-tritium neutron generator specified and measured to be capable of producing 14 MeV neutrons at rates of up to 8.2$\times$10$^8$~neutrons/s. The generator is enclosed in a carbon-steel  32$^{\prime\prime}$ diameter, 23.62$^{\prime\prime}$ high carbon-steel cylinder that moderates the mean neutron energy to under 500 keV. This, in turn, is encased in 5\%-borated polyethylene such that the entire assembly is a 48$^{\prime\prime}$$\times$48$^{\prime\prime}$ box that is 30$^{\prime\prime}$ tall. For radiographic applications, a narrow, tapered channel in the steel and polyethylene allows 14 MeV neutrons to stream directly from the generator to a test object. Its collimating capability is demonstrated by measuring the neutron flux profile. In the moderated mode of operation, the generator is fully enclosed in the steel, but a large section of the polyethylene is removed, providing a flux of sub-MeV neutrons from a wide range of angles. Neutron angular and spectral measurements using both a nested neutron spectrometer and a commercial liquid scintillator coupled with a $^3$He detector show the expected softer neutron spectrum in moderated mode in good agreement with MCNP6 calculations. The gamma-ray spectrum from Excalibur is also in good agreement with MCNP modeling. Based on these findings, the future application of Excalibur in its two configurations is discussed.

\end{abstract}

\begin{keyword}
Nuclear warhead verification \sep
source characterization \sep
neutron radiography \sep
fissile material detection
\end{keyword}

\end{frontmatter}

%\linenumbers

% **********************************************************************
% **********************************************************************
\section{Introduction}
\label{sec:introduction}

A reliable, well-characterized neutron source is important for testing and validating a variety of potential warhead verification schemes, including radiographic imaging \cite{philippe} and the detection of neutrons from induced fission in fissile material \cite{yan}. The Experiment for Calibration with Uranium (Excalibur) is a neutron source that has been developed to serve these two purposes. It can be used in two distinct modes: a collimated mode that produces a fan-shaped beam of 14 MeV neutrons for radiography; and a moderated mode that produces a uniform broad flux of sub-MeV neutrons to induce fission events in the fissile material present in the target \cite{yan, hepler}. 

% ****************************
\begin{figure}[htb]
    \centering \sf
    \includegraphics[width =\textwidth]{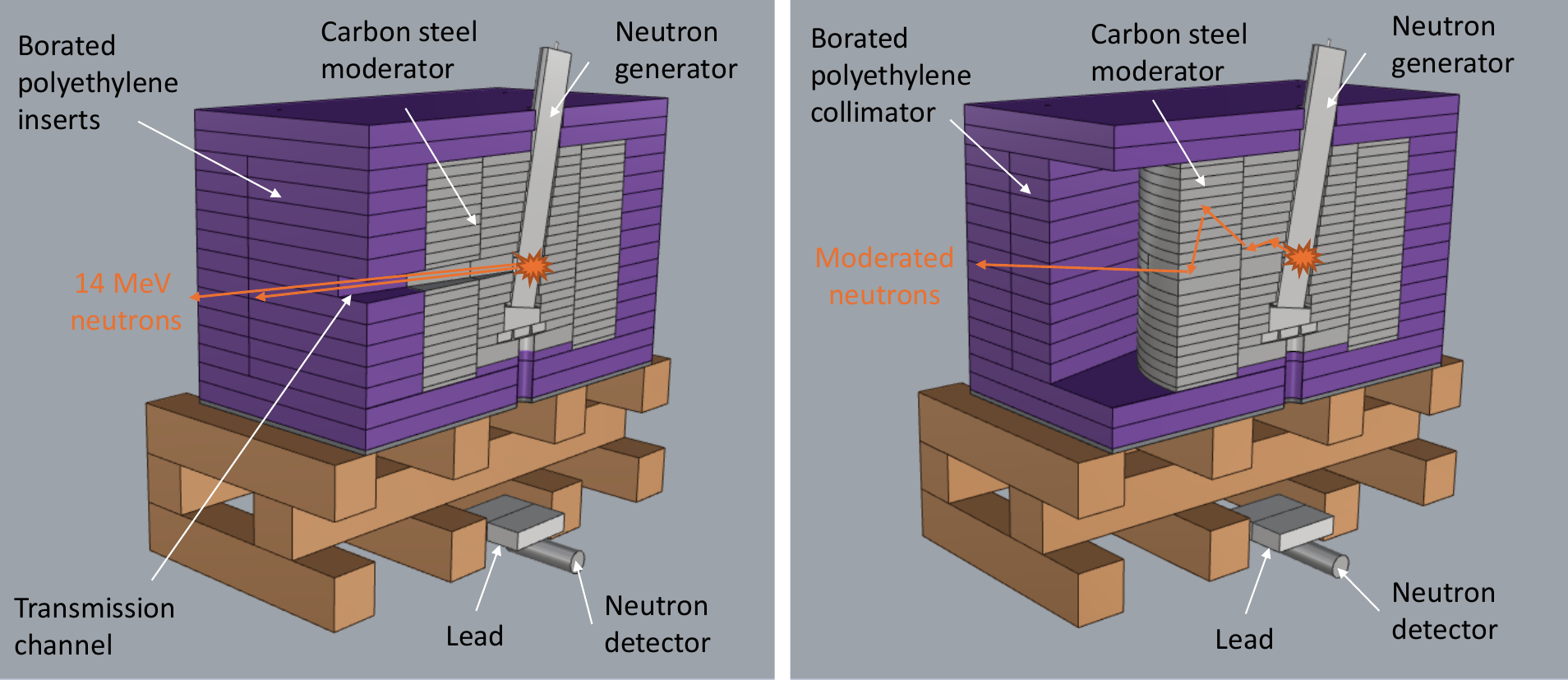}
    \caption{Shown on the left is a cutaway view of half of Excalibur in collimated mode. The DT generator sits at the center of a carbon-steel cylinder and is inclined by $10^\circ$ in order to operate with vertically uniform flux across the exit aperture. A horizontal aperture with an opening angle of $17.5^\circ$ allows a fan-shaped beam of 14~MeV neutrons to be emitted. An Eljen-410 ZnS fast neutron detector below the DT generator monitors neutron emission. Shown on the right is a cutaway view of moderated mode. Few 14~MeV neutrons escape, while a broad flux of sub-MeV neutrons reaches the test object and drives the production of fission neutrons if fissile material is present.}
    \label{fig:EXCALIBUR}
\end{figure}
% ****************************

\begin{figure}[htb]
    \centering \sf
   \includegraphics[width = \textwidth]{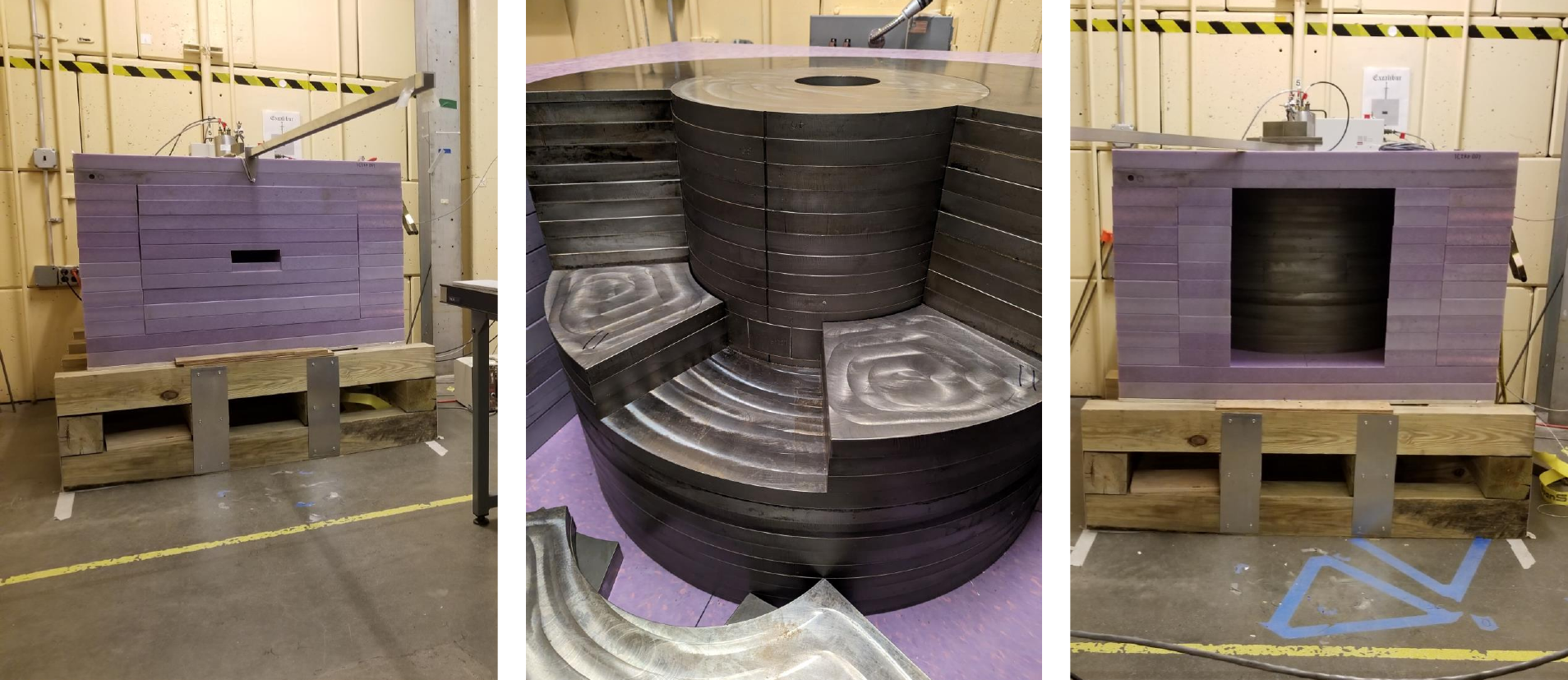}
    \caption{Shown on the left is the as-built appearance of Excalibur in collimated mode. Shown in the middle is Excalibur during reconfiguration between modes. The top layers of purple polyethylene and the DT generator have been removed to allow access to the carbon-steel cylinder components. Shown on the right is the as-built appearance of Excalibur in moderated mode.}
    \label{fig:Reconfig}
\end{figure}

The heart of the Excalibur source is a Thermo Fisher P-385 deuterium-tritium (DT) neutron generator that produces up to $4.5\times10^8$~neutrons/s with the A3082 tube and $8.2\times10^8$~neutrons/s with the newer A3083 tube as specified and measured in \cite{jeon_dt}. In both cases, neutron production scales linearly with beam current and follows a power law with an exponent of about 3.5 as the accelerating voltage increases, i.e., $\dot{n}(V) \propto V^{3.5}$, as predicted by TRIM-based simulations \cite{jeon_dt}. As illustrated in Figure~\ref{fig:EXCALIBUR}, the DT generator is placed in the center of a carbon-steel cylinder made of individual disks---like a sword in a stone, the source of its name, Excalibur, coined by Senior Health Physics Technician, Darren Thompson. The generator is angled backwards at 10$^\circ$ to avoid a steep flux gradient observed in the target plane due to a heat sink attached to the tritiated target \cite{hepler}.
The construction method allows the 32-inch-diameter, 60-cm-high cylinder to be reconfigured between modes in under one hour, without the need for special tools or lifting equipment (Figure~\ref{fig:Reconfig}). The steel cylinder is encased, except for the desired neutron exit aperture, in a block of 5\%-borated polyethylene that further moderates and absorbs neutrons that make it through the steel. In collimated mode (Figure~\ref{fig:EXCALIBUR}, left and Figure~\ref{fig:Reconfig}, left), steel and polyethylene are removed to create a tapered channel so that 14 MeV neutrons can stream directly to a target. In moderated mode (Figure~\ref{fig:EXCALIBUR}, right and Figure~\ref{fig:Reconfig}, right), the steel cylinder is complete, but the majority of the polyethylene on one face of Excalibur is removed so that a test object can face the full diffuse flux from the cylinder. The geometry and material for the collimator and moderator have been optimized for the desired neutron energy spectrum, maximizing the number of neutrons below 1~MeV so that these source neutrons can be distinguished from higher-energy neutrons originating from fission events in the target \cite{hepler}.

% ****************************
% ****************************

The Excalibur neutron source must be well-characterized in order to be useful for exploring and validating proposed methods that depend on a reliable known neutron flux distribution. To this end, in this paper we present spectrally-resolved neutron flux measurements as a function of the horizontal angle relative to the neutron bream axis in two different modes and provide a detailed discussion of their neutron spectra. We further explore the gamma-ray spectrum from Excalibur. The results are shown to be in good agreement with MCNP6 simulations of Excalibur.

% **********************************************************************
% **********************************************************************

\section{Methods}

We used three different types of neutron detectors to fully characterize Excalibur as part of this study. Two of these detectors are part of the BTI N-Probe neutron spectrometer, which combines a He-3 counter for detecting neutrons up to 0.8~MeV and a liquid scintillator for detecting neutrons with energies from 0.8~MeV to 20~MeV \cite{ing}. The pulse-height spectrum from the liquid scintillator and the boron-shielded He-3 counter is processed using a proprietary unfolding software. The resulting unfolded spectrum from the N-Probe offers five energy bins below 0.8~MeV and 48 bins between 0.8~MeV and 20~MeV, enabling an unfolded neutron spectrum across a wide energy range \cite{koslowsky}, though with substantial uncertainties in some energy regions. In this paper, the bins below 0.8 MeV were summed because the error bars for each energy bin were high due to the low He-3 counts, particularly at the higher energies. The BTI N-Probe was calibrated by the manufacturer with a Cf-252 source, resulting in an uncertainty in dose rate of about 13\% at the 95\% confidence level. The calibration was performed using standards traceable to National Institute of Standards and Technology (NIST).

The third detector used for this study is a nested neutron spectrometer (NNS) by Detec \cite{detec}, which offers better resolution for low-energy neutron than the N-Probe. The NNS comprises seven polyethylene cylindrical shells surrounding a He-3 detector, and it is capable of determining neutron energies from thermal energies to 20~MeV. The N-Probe provides neutron spectra with linearly spaced energy bins, each 0.4 MeV wide, the NNS provides a neutron spectra with logarithmically spaced energy bins, with five bins per decade. The NNS is therefore better suited for measurements on thermal and epithermal neutrons. The NNS has been calibrated with an AmBe standard source of the Ionizing Radiation Standards Laboratory at the National research Council of Canada (IRS-NRC). The detector calibration tests were conducted in a low-scatter facility and the neutron output of the AmBe source, with a 2\%  uncertainty, is traceable to NIST \cite{detec}. Considering the unfolding process, which fills 52 bins with only 7 independent measurements, the detailed energy spectrum should be viewed with some caution.

\begin{figure}[!htb]
    \centering \sf
   \includegraphics[width = \textwidth]{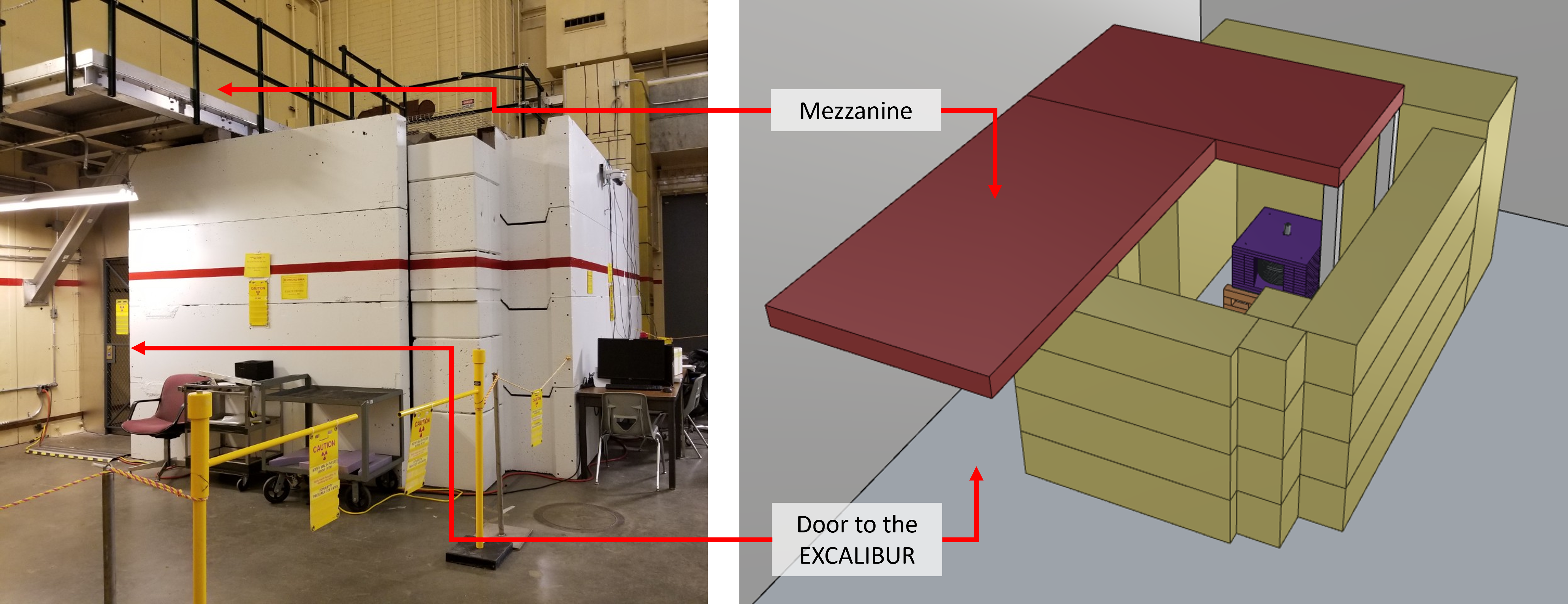}
    \caption{Excalibur in Tokamak Fusion Test Reactor (TFTR) test cell at Princeton Plasma Physics Laboratory (PPPL). The left figure is a photo showing the concrete shield walls around Excalibur, while the right figure is a view from above, generated using a Rhinoceros 3D model, depicting the real setting and showing the Excalibur inside the concrete walls.}
    \label{fig:EXCALIBUR_TFTR}
\end{figure}

To complement measurements, we also simulated the experimental setup using MCNP6 Monte Carlo calculations \cite{werner}. The MCNP6 model includes a detailed model of the neutron generator provided by the manufacturer, a faithful reproduction of Excalibur, and the environment where the experiments were conducted. The experimental setup is located in the well-shielded Tokamak Fusion Test Reactor (TFTR) test cell at the Princeton Plasma Physics Laboratory (PPPL). The test cell measures 46~m in width, 35~m in length, and 16.5 m in height. The concrete wall and roof are 1.25 meters thick, while the floor is 1.5 meters thick. Excalibur is located in a corner of the test cell inside an additional arrangement of thick concrete shield walls. The entire test cell was modeled in MCNP6 when thermal neutrons from room return were expected to affect the measurements, such as those with the He-3 bare detector. Otherwise, only the immediate area containing the Excalibur neutron source, surrounded by a 80~cm thick concrete shielding wall, was modeled as shown in Figure~\ref{fig:EXCALIBUR_TFTR}.

% This area measures 7.31 m (W) x 7.61 m (L) x 5.14 m (H). 

% ****************************
% ****************************

% **********************************************************************
% **********************************************************************

\section{Collimated Mode}

In collimated mode, a tapered neutron exit aperture is created in the steel cylinder and polyethylene block with a total opening angle of 17.5$^\circ$ and a height of 2 inches. To characterize the performance of the system, we spatially scan the neutron flux with the N-Probe. Since we expect direct 14 MeV neutrons in collimated mode, neutron flux in the energy above 10 MeV is profiled vs. horizontal angle from the beam axis. We also compare these results against MCNP6 simulations and the idealized notion of single-flight (or uncollided) ballistic neutron trajectories from a point source.

\begin{figure}[!htb]
    \centering \sf
    \includegraphics[width=110mm]{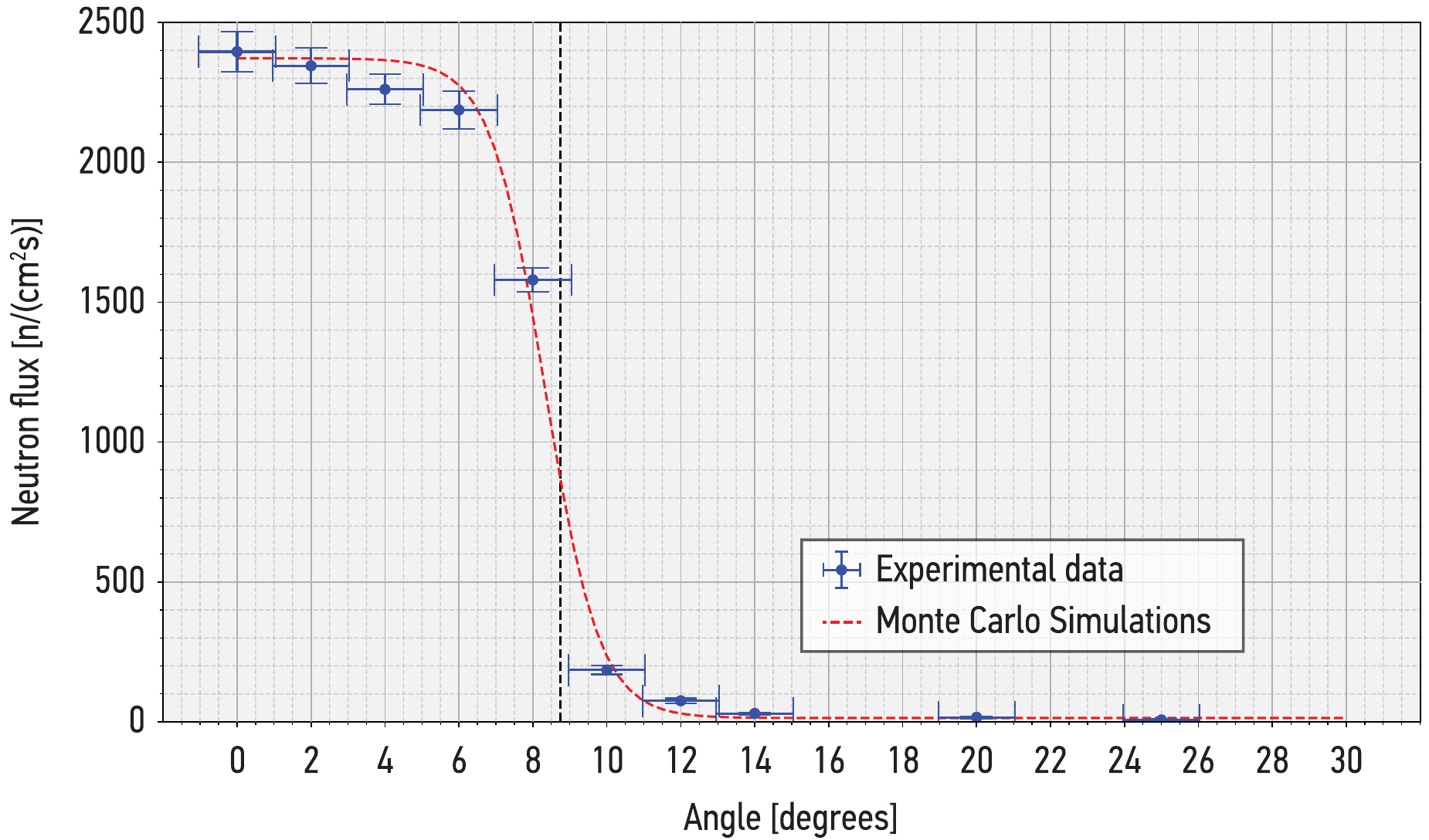}
    \caption{Neutron flux as a function of horizontal angle from the beam axis in Excalibur collimated mode. The detector is located at 1.4 m from the source and neutron flux in the unfolded spectrum was integrated over the energy range of 10--20 MeV. Simulated results are based on a series of MCNP6 calculations (for small increments in observation angle) and show the neutron flux integrated over the same energy range. The dashed vertical line indicates the ideal edge of the aperture at 8.75$^\circ$ off-axis. The high-energy neutron flux has dropped by more than 95\% for an observation angle of 12$^\circ$ off-axis.}
    \label{fig:CollimatedScan}
\end{figure}

During the measurements, the N-Probe was first placed along the neutron beam axis at a fixed distance of 1.4 m from the source (A3083 tube at a current of 70~$\mu$A and a voltage of 130~kV) and then rotated horizontally in two-degree increments from 0$^\circ$ to 14$^\circ$ to one side. The scintillator has a 2-inch diameter, represented as horizontal error bars in Figure~\ref{fig:CollimatedScan}, while the vertical error bars are derived from the standard deviation of the mean of ten separate two-minute measurements. Left-right symmetry was assumed. Above 10 MeV, experimental results and simulations show that the neutron flux is nearly flat from the beam axis to 5$^\circ$ off-axis, falling to somewhat less than half its on-axis value at the edge of the ideal ballistic aperture at 8.75$^\circ$ off-axis.

% ****************************
% ****************************

\begin{figure}[!htb]
    \centering \sf
    \includegraphics[width = 0.9\textwidth]{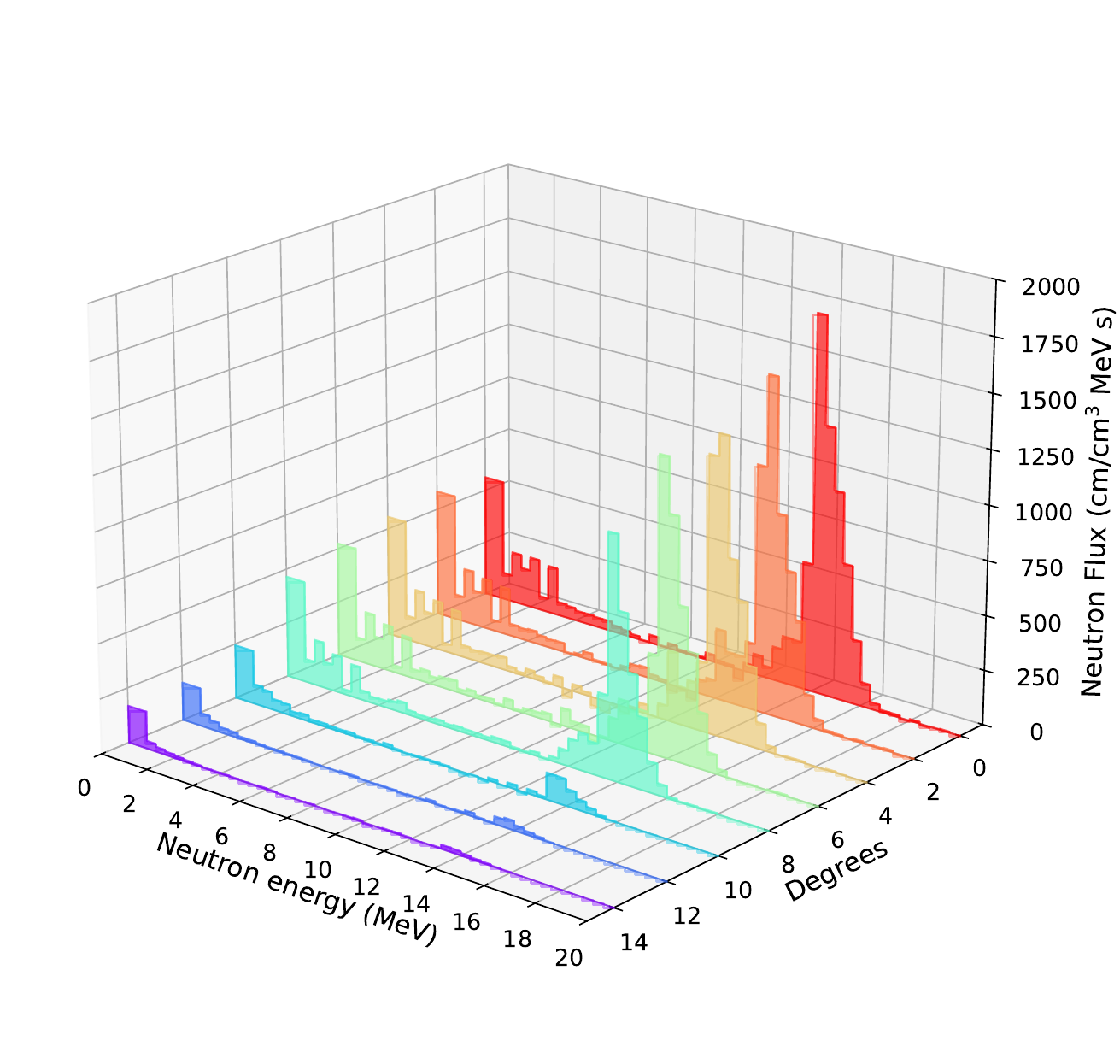}
    \caption{Neutron spectra of Excalibur in collimated mode. These spectra have been acquired with the N-Probe and unfolded with the instrument's provided algorithm. The bins below 0.8 MeV were summed because the error bars for each energy bin were high due to the low He-3 counts.}
    \label{fig:CollimatedSpec}
\end{figure}

We also examined the variation in neutron energy as a function of the off-axis angle (Figure~\ref{fig:CollimatedSpec}). As expected, the primary beam contains mostly 14-MeV neutrons, but with a relevant contribution below 2 MeV. For angles greater than 10--12$^\circ$, in the shadow of the collimator, the absolute neutron flux drops significantly for all energies, and the population of low-energy neutrons becomes significant compared to the high-energy flux.

% ****************************
% ****************************

\clearpage
% **********************************************************************

\section{Moderated Mode}

In moderated mode, we aim to use sub-MeV source neutrons to drive fission events in an inspected object. The resulting fission neutrons, in the 1--4 MeV range, would then be detected. A key aspect of this mode is moderating the source neutrons to below 1 MeV, enabling detectors with an energy threshold to distinguish them from fission neutrons. This configuration of Excalibur has been optimized such that the sub-MeV flux is as high as possible, both in absolute terms and in relative terms compared to the higher-energy flux. To achieve this goal, the DT generator is completely enclosed by the carbon-steel cylinder. Most of one face of the borated polyethylene outer layer has been removed, which enables a broad flux of sub-MeV neutrons to emerge from the steel surface. In moderated mode, a broad profile of uncollimated sub-MeV neutrons is expected and required to provide adequate flux at a test object.

\begin{figure}[!htb]
    \centering \sf
    \includegraphics[width=0.9\textwidth]{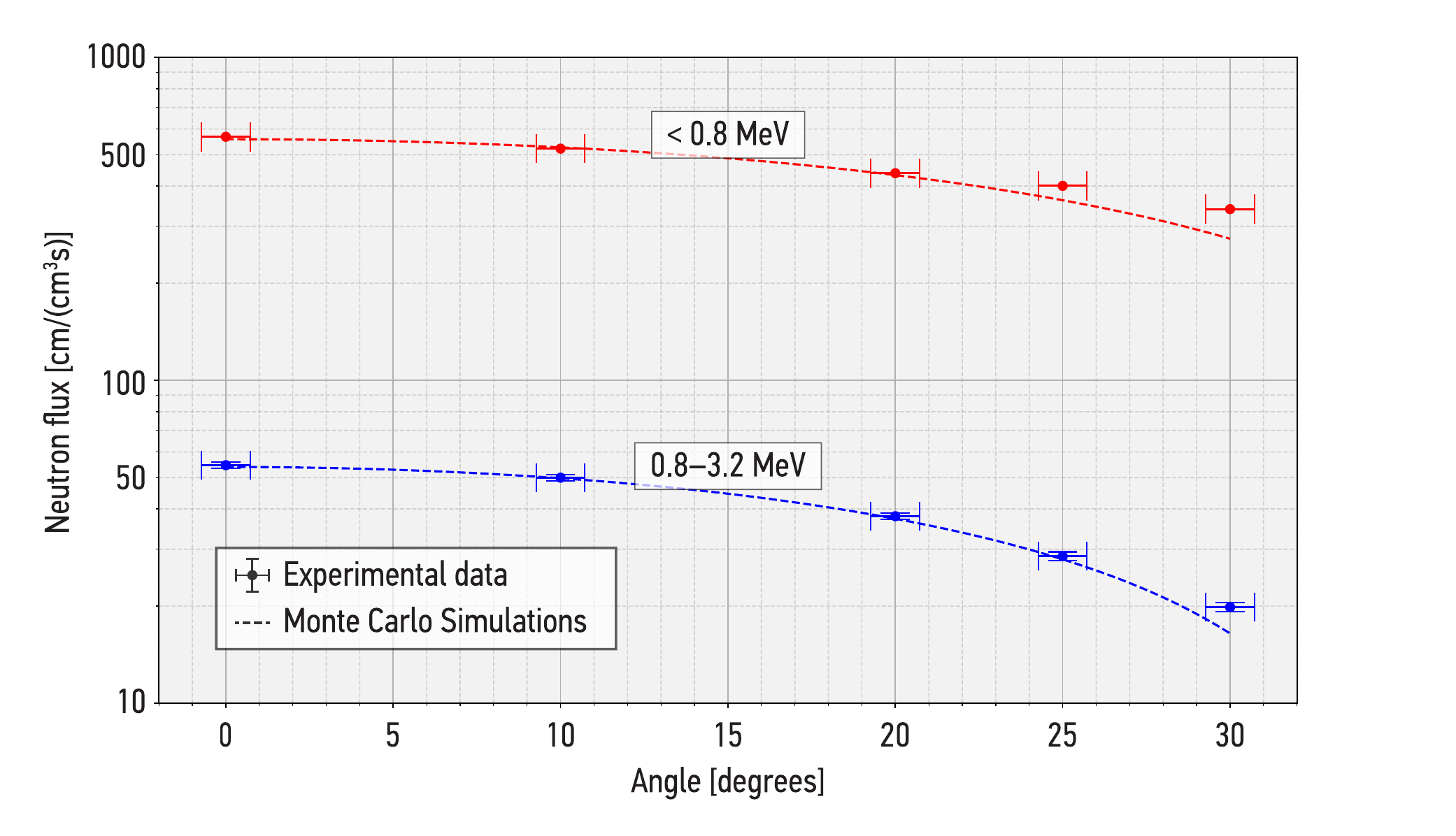}
    \caption{Neutron flux as a function of horizontal angle from the beam axis in Excalibur moderated mode. The detector is located at 2 m from the source and neutron flux was estimated across the energy ranges of 0.0--0.8 MeV (sub-MeV neutrons) and 0.8--3.2 MeV. The spatial profile for 0.0--0.8 MeV was derived by summing over from an unfolded spectrum acquired by the N-Probe He-3 detector, while the profile for 0.8--3.2 MeV was obtained from the spectrum by the liquid scintillator. The MCNP6 model was used to calculate the neutron flux accordingly. Simulated results are based on a series of MCNP6 calculations for small increments in observation angle. The neutron flux in the 10--20 MeV energy range, with a significant error bar, is two orders of magnitude lower than the flux below 0.8 MeV.}
    \label{fig:ModeratedScan}
\end{figure}

Figure~\ref{fig:ModeratedScan} shows experimental and simulated neutron flux levels as the measurement angle varies from 0 degrees to 30 degrees off axis. Five two-minute measurements were taken at each location at distance of 2 m from the source (A3082 tube at a current of 40~$\mu$A and a voltage of 130~kV.) We plot the neutron flux below 0.8~MeV from the He-3 detector as well as the neutron flux in 0.8--3.2~MeV from the liquid scintillator.

In contrast to the rather sharp drop-off of neutron flux with angle for the collimated mode in Figure~\ref{fig:CollimatedScan}, the low-energy neutron flux in moderated mode case varies smoothly from its on-axis value to nearly half that at an angle slightly over 25 degrees. Experimentally measured neutron fluxes below 0.8~MeV and 0.8--3.2~MeV are consistent with those computed with MCNP6.

% **********************************************************************

% ****************************
% ****************************

\newpage

\begin{figure}[!htb]
    \centering \sf
    \includegraphics[width = 0.8\textwidth]{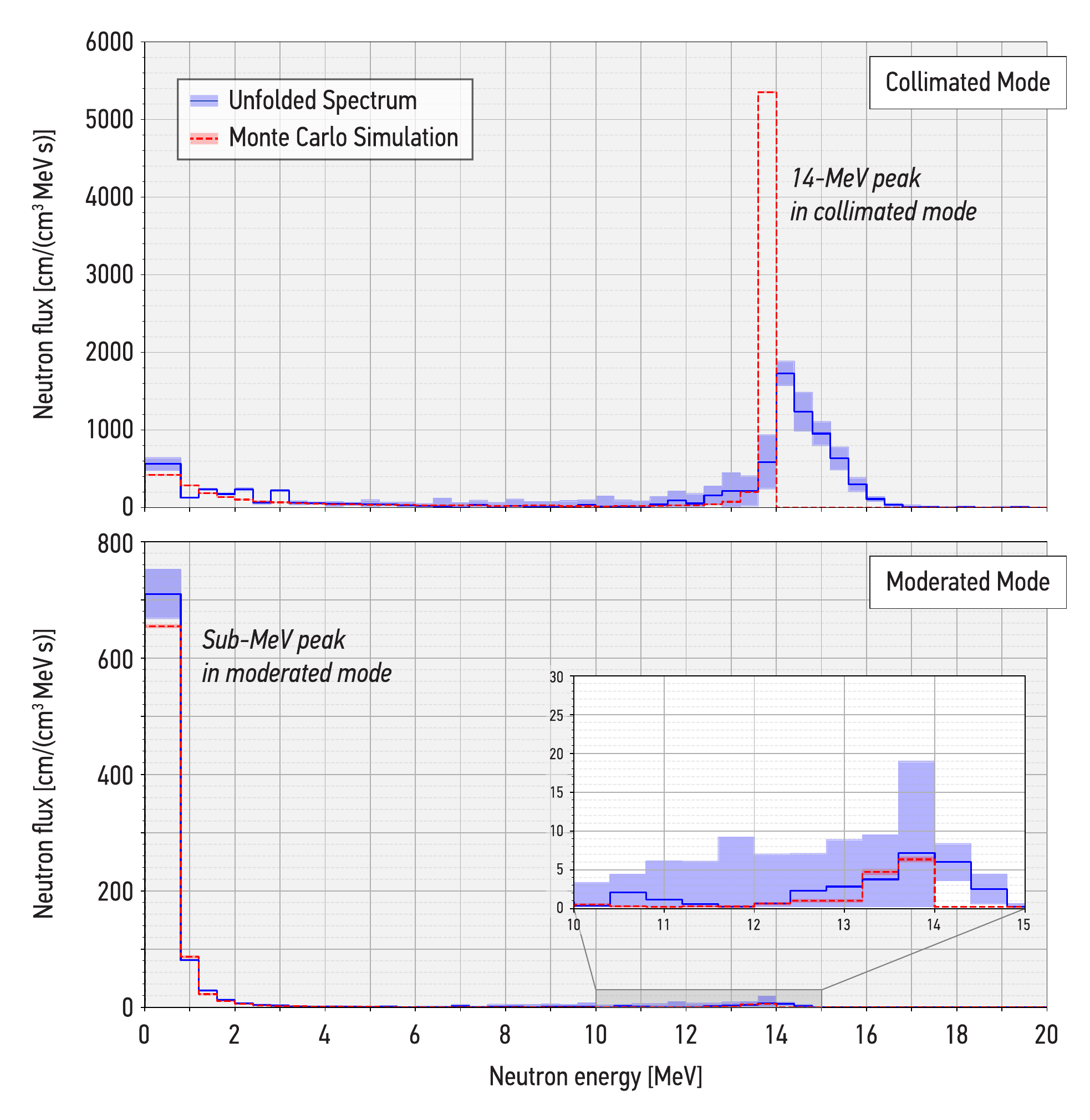}
    \caption{On-axis comparison of collimated and moderated mode. 
    Neutron flux spectra in collimated mode (top panel) are measured with the N-Probe and simulated in MCNP6 at a distance of 1.4~meters from the source (A3083 tube.) Measurements are for a current of 70~$\mu$A and a voltage of 130~kV. Neutron flux spectra in moderated mode (bottom panel) are measured and simulated at a distance of 2~meters from the source (A3082 tube); for operational reasons, these measurements used a reduced current of 40~$\mu$A. Simulated spectra use a scaling factor that has been determined using data from previous experiments characterizing the output of the neutron generator as a function of operating current and voltage \cite{jeon_dt}. The shaded areas indicate the estimated standard deviations; in the case of the N-Probe measurements, these deviations were estimated based on ten two-minute measurements as previously.}
    \label{fig:col_mod_spec}
\end{figure}

Experimental and MCNP6 on-axis neutron spectra are shown in Figure~\ref{fig:col_mod_spec} for both collimated and moderated modes. To facilitate a direct comparison, the 53 energy bins of the N-Probe were also also adopted for the Monte Carlo simulations. Overall, the data are in good agreement, both in terms of shape and absolute neutron yield. This is notable considering the finite energy resolution of the N-Probe and the uncertainties involved in the spectrum unfolding process.

Unfolded and simulated spectra confirm a strong 14-MeV peak in neutron flux when Excalibur operates in collimated mode, and a pronounced sub-MeV peak in moderated mode. In collimated mode, neutrons within 10--20 MeV energy range account for 70\% of the total neutron flux, whereas in moderated mode, neutrons below 0.8 MeV make up 89\% of total flux. This shift in the spectrum to a softer one in moderated mode demonstrates the effectiveness of the carbon steel moderator. Additionally, in terms of absolute flux values across all neutron energies, the collimated mode produces 1.6 times greater neutron flux than the moderated mode.

The unfolded spectra produced with the N-Probe also show, however, significant errors in regions where flux levels are low (4--12 MeV), leading to poor reproducibility of the measurements there. Similarly, neutron energies above 14 MeV reported in collimated mode are an artifact of the unfolding algorithm; when the counts in these bins are combined, they match the counts in the 14-MeV bin predicted by the MCNP6 simulations. Despite these limitations, the regions of interest, i.e., 0.8--3.2 MeV in moderated mode and 10--20 MeV in collimated mode, show good agreement between the experimental results and MCNP6 calculations.

% ****************************

% ****************************

\clearpage

In moderated mode, the neutron flux below 0.8 MeV was 4.5 times higher than in the collimated mode, for the same source rate. The ratio of neutron flux in the 0--0.8 MeV range to that in 10--20 MeV range was 274 times higher in moderated mode than in collimated mode. While the N-Probe spectra and MCNP6 calculations provide good insight into the moderating power of the Excalibur moderated mode, there are still limitations in fully understanding the neutron profile below 1 MeV due to the low resolution of energy bins.

\begin{figure}[!htb]
    \centering \sf
    \includegraphics[width=110mm]{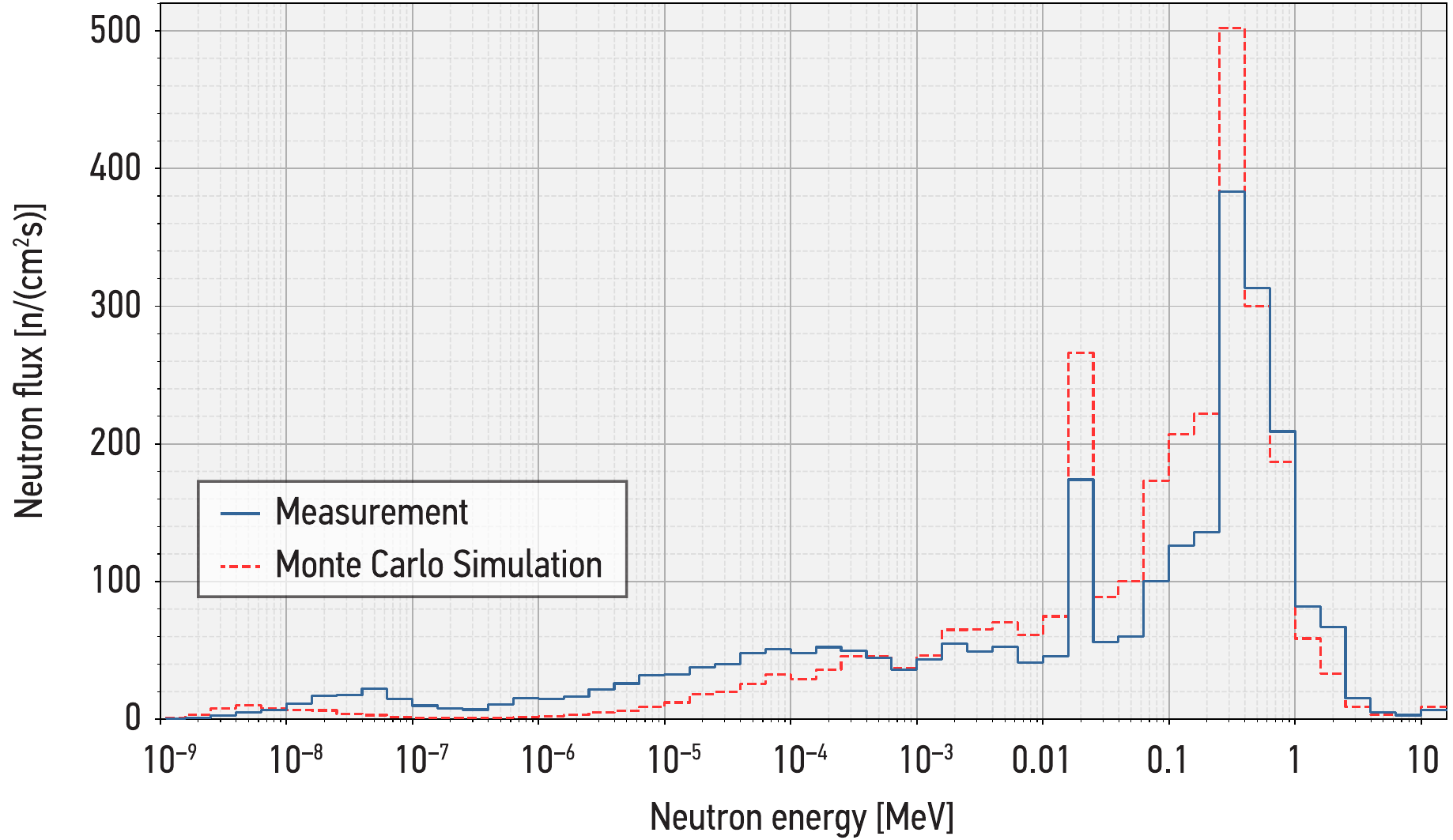}\\
    \caption{Experimental and simulated neutron spectrum in moderated mode. Neutron flux spectra in moderated mode as measured by the Nested Neutron Spectrometer and as calculated by MCNP6 at a distance of 1 meters from the source (A3082 tube.) Measurements are for a current of 60 $\mu$A and the voltage of 100 kV \cite{hepler}; for the simulated spectra, the scaling factor has been determined using data from on previous experiments characterizing the current and voltage dependencies \cite{jeon_dt}. The NNS unfolds the spectrum using an a priori guess spectrum calculated with MCNP.}
    \label{fig:nns}
\end{figure}

In this context, the NNS was used to acquire a neutron spectrum more detailed in the thermal and epithermal region. The eight neutron count rates were obtained with a minimum of 12,000 counts and then unfolded using the MLEM algorithm to derive the neutron energy spectrum. Corresponding count rates were calculated with MCNP6, with each individual shell models and unfolded using the same algorithm. Considering the limitations of the MLEM algorithm for this under-determined problem \cite{montgomery}, the NNS measurements show good agreement with the MCNP6 calculations, as illustrated in Figure~\ref{fig:nns}. It implies a neutron energy spectrum with a peak in the range of 250--400 keV. MCNP6 has lower neutron flux in the energy below 1 keV than the measurement presumably due to inadequate modeling of room-return neutrons. The peak at 15.9--25.1 keV comes from the low neutron scattering cross-section of $^{56}Fe$ in energy around 24 keV. From this, we can state that the moderated mode succeeded in moderating neutrons to the energy of our interest. This is consistent with the requirements for the planned fissile material detection experiments.

% ****************************
% ****************************

\section{Gamma-ray Spectrum}

\begin{figure}[!htb]
    \centering \sf
    \includegraphics[width=110mm]{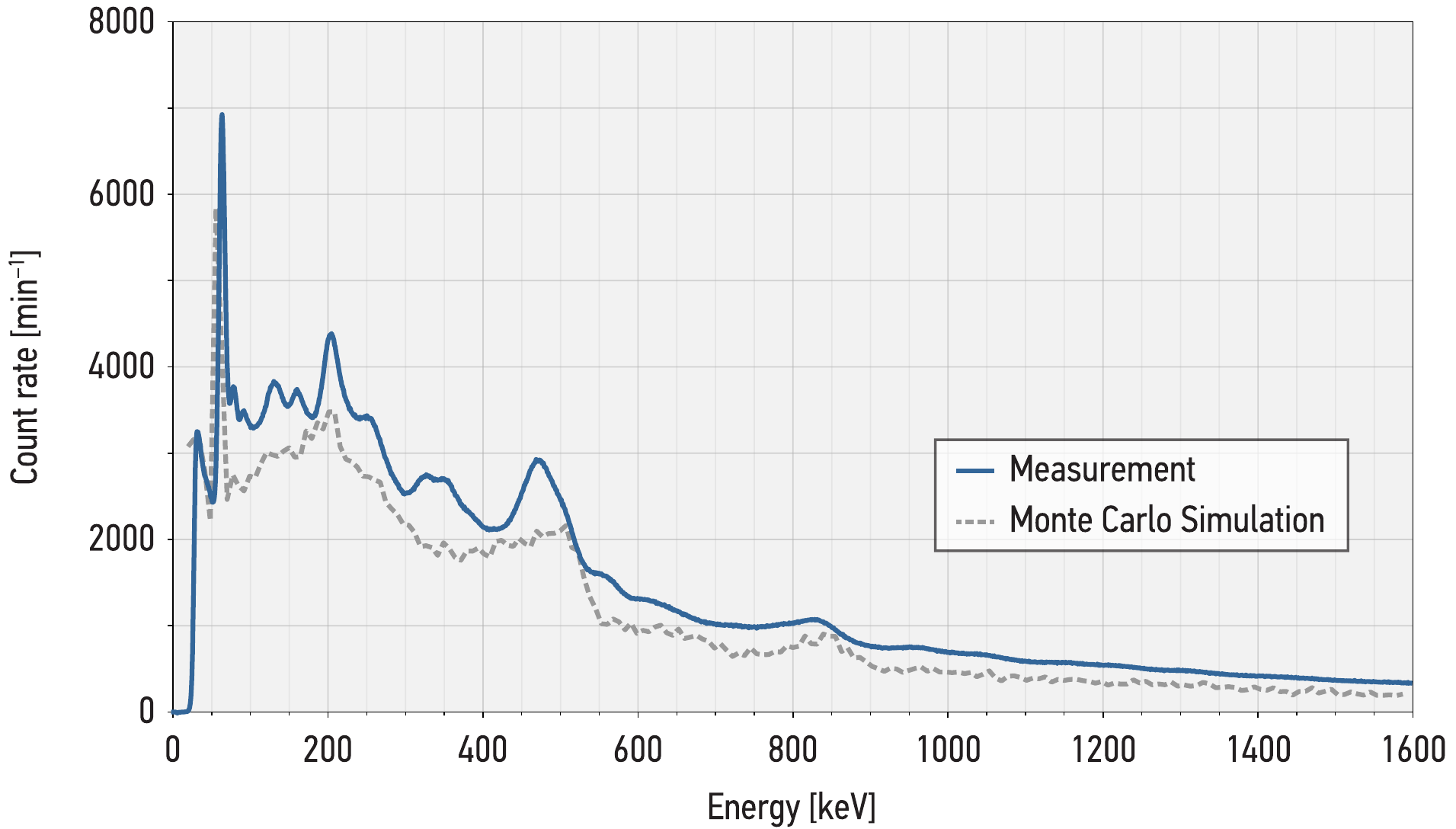}
    \caption{Gamma ray spectrum from Excalibur in collimated mode, measured by a NaI(Tl) scintillator and simulated with MCNP6. During the measurement, we placed the detector at a distance of 2 meters from the source running at a current of 70~$\mu$A and a voltage of 130~kV. For the calculation, we took the source rate in accordance with previous experiments on the current and voltage dependence \cite{jeon_dt}.}
    \label{fig:gamma_spec}
\end{figure}

The neutron detectors we have been using are essentially blind to gamma-rays: the N-Probe has a gamma rejection capability of approximately 1000:1 at 20$\,^{\circ}$C with its pulse-shape discrimination circuitry \cite{ing, nprobe}, and the $^3$He counters (both in N-Probe and NNS) are known to interact with gamma radiation very weakly. Still, it is useful to characterize the gamma flux from Excalibur, as other imaging detectors are considered. To do so, we acquired a 3-minute gamma-ray spectrum with a 2-inch NaI detector placed at a distance of 2 meters from the Excalibur in collimated mode. The neutron source (A3082 tube) was set at the current of 70 $\mu$A and voltage of 130 kV. For the spectrum simulated in MCNP6, we used the activation control (ACT) card to calculate gamma rays produced by neutron activation and the particle track (PTRAC) card to identify the isotopes and the energies of the specific gamma rays they generated. Results are shown in Figure~\ref{fig:gamma_spec}. The prominent peaks at 57 keV and 203 keV are associated with 14-MeV neutron inelastic scattering on $^{127}I$ in the NaI(Tl) detector \cite{simakov}.

% ****************************
% ****************************

Other than the gamma rays produced from the direct interaction between the NaI detector and 14 MeV neutrons, the MCNP6 PTRAC analysis indicated that boron-10 and iron-56 were the isotopes emitting the most gamma rays. Since the neutron generator is surrounded by a thick carbon steel cylinder in a block of 5\%-borated polyethylene, the highest peak above 300 keV is located at 478 keV in the MCNP6 spectrum, which comes from boron-10 interaction with 14 MeV neutrons. The next highest peak is the annihilation peak at 511 keV, followed by 847 keV. The 847 keV gamma peak is associated with inelastic scattering and $(n,2n)$ reactions on Fe-56 \cite{simakov}. 
% **********************************************************************

\section{Discussion}

There are many possible use cases for Excalibur based on its unique characteristics discussed. here. 
First of all, our group used the Excalibur neutron source to develop a radiation detection system for application of zero-knowledge protocol (ZKP) imaging \cite{glaser}. We will continue to work on ZKP radiography in the Excalibur collimated mode with superheated droplet detectors \cite{jeon_bubble1, jeon_bubble2}. We tried a ZnS scintillator plate, coupled with photographic film as an alternative to the droplet detectors \cite{jeon_photo} but the signal level was low. We plan to use a EMCCD camera to produce radiographs which combine or subtract a complementary signal to achieve the goal of zero-knowledge \cite{goldston}.

While the collimated mode with 14 MeV neutrons is suitable for neutron radiography, 14 MeV neutron radiography is not sensitive to $^{235}$U vs.$^{238}$U and $^{239}$Pu vs.$^{240}$Pu. Therefore, one can use the moderated mode with sub-MeV neutrons to detect fissile isotopes in the nuclear material. If the test object contains fissionable materials, and the detectors are sensitive only to supra-MeV neutrons, then the neutrons from Excalibur's moderated mode should be suitable to distinguish fission neutrons generated in the test object from source neutrons. To maximize the signal to noise ratio and shield the detectors from source neutrons for this application, further optimization of the Excalibur moderated mode design was completed \cite{hepler, gilbert} and the additional moderating and shielding system is in-place at PPPL. Beside the zero-knowledge application, one can use the Excalibur neutron source for other various measurements such as neutron die-away analysis, beta-delayed neutrons and neutron multiplicity.

An advantage of the Excalibur for these applications is that the system can be reconfigured between its two modes in under one hour, providing flexibility in experimental operations. A ZnS fast neutron detector (EJ-410) is located under the generator to monitor source performance and it showed us the system reproducibility better than 1\% fluctuation. Excalibur is located in a shielded enclosure, comprised of thick shield walls, resulting in very low radiation dose to personnel.

% **********************************************************************
% **********************************************************************

\section{Conclusion}

The Excalibur neutron source is a flexible and reliable system for testing warhead verification concepts that require well-characterized, steady, reliable 14~MeV neutron beams and broad sub-MeV neutron fluxes. The core component of the source, a DT neutron generator, has previously been well characterized in terms of its operating condition, neutron flux and gamma-ray emission. A BTI N-Probe has been used to characterize the spatial and energy distribution emitted from the source in two modes: collimated and moderated. The high energy neutrons in the collimated mode follow the tapered shape of the aperture, while the sub-MeV neutrons in moderated mode are emitted in a broad uniform flux, both consistent with MCNP6 calculations. The moderating capability of the Excalibur was confirmed by N-Probe as well as NNS. The Excalibur is expected to not only serve its purpose for zero-knowledge verification but also be versatile for other neutron experiments which require 14 MeV high energy neutrons or sub-MeV neutrons. We expect the source characterization study in this paper will be conducive to the future use of the Excalibur.

% **********************************************************************

\section*{Acknowledgements}

This work was supported by the U.S. Department of Energy under contract number DE-AC02-09CH11466. The United States Government retains a non-exclusive, paid-up, irrevocable, world-wide license to publish or reproduce the published form of this manuscript, or allow others to do so, for United States Government purposes. Authors gratefully acknowledge the contribution of all the PPPL staffs for reconfiguring the Excalibur---Matthew Beer, Robert Bongiovanni, Andrew Carpe, Russell Gutter III, Robert Hitchner, John Horner, Kevin Purdy, John Rayment, and Darren Thompson.
% **********************************************************************

\end{document}